\documentclass[twocolumn]{aastex62}
\usepackage{natbib}
\usepackage{threeparttable}
\usepackage{amssymb}
\usepackage{graphicx}
\usepackage{CJKutf8}
\usepackage{textcomp}

\shorttitle{Recently quenched galaxies}
\shortauthors{Wu et al.}

\begin{document}

\title{The colors and sizes of recently quenched galaxies: a result of compact starburst before quenching}

\author{Po-Feng Wu \begin{CJK*}{UTF8}{bkai}(吳柏鋒)\end{CJK*}}
\altaffiliation{EACOA Fellow}
\affiliation{National Astronomical Observatory of Japan, Osawa 2-21-1, Mitaka, Tokyo 181-8588, Japan}
\affiliation{Max-Planck-Institut f\"{u}r Astronomie, K\"{o}nigstuhl 17, D-69117, Heidelberg, Germany}

\author{Arjen van der Wel}
\affiliation{Sterrenkundig Observatorium, Universiteit Gent, Krijgslaan 281 S9, B-9000 Gent, Belgium}
\affiliation{Max-Planck-Institut f\"{u}r Astronomie, K\"{o}nigstuhl 17, D-69117, Heidelberg, Germany}

\author{Rachel Bezanson}
\affiliation{University of Pittsburgh, Department of Physics and Astronomy, 100 Allen Hall, 3941 O'Hara St, Pittsburgh PA 15260, USA}

\author{Anna Gallazzi}
\affiliation{INAF-Osservatorio Astrofisico di Arcetri, Largo Enrico Fermi 5, I-50125 Firenze, Italy}

\author{Camilla Pacifici}
\affiliation{Space Telescope Science Institute, 3700 San Martin Drive, Baltimore, MD 21218, USA}

\author{Caroline M. S. Straatman}
\affiliation{Sterrenkundig Observatorium, Universiteit Gent, Krijgslaan 281 S9, B-9000 Gent, Belgium}

\author{Ivana Bari\v{s}i\'{c}}
\affiliation{Max-Planck-Institut f\"{u}r Astronomie, K\"{o}nigstuhl 17, D-69117, Heidelberg, Germany}

\author{Eric F. Bell}
\affiliation{Department of Astronomy, University of Michigan, 1085 South University Avenue, Ann Arbor, MI 48109-1107, USA}

\author{Priscilla Chauke}
\affiliation{Max-Planck-Institut f\"{u}r Astronomie, K\"{o}nigstuhl 17, D-69117, Heidelberg, Germany}

\author{Francesco D’Eugenio}
\affiliation{Sterrenkundig Observatorium, Universiteit Gent, Krijgslaan 281 S9, B-9000 Gent, Belgium}

\author{Marijn Franx}
\affiliation{Leiden Observatory, Leiden University, PO Box 9513, 2300 RA Leiden, The Netherlands}

\author{Adam Muzzin}
\affiliation{Department of Physics and Astronomy, York University, 4700 Keele St., Toronto, Ontario, M3J 1P3, Canada}

\author{David Sobral}
\affiliation{Physics Department, Lancaster University, Lancaster LA1 4YB, UK}
\affiliation{Leiden Observatory, Leiden University, PO Box 9513, 2300 RA Leiden, The Netherlands}

\author{Josha van Houdt}
\affiliation{Max-Planck-Institut f\"{u}r Astronomie, K\"{o}nigstuhl 17, D-69117, Heidelberg, Germany}

\begin{abstract}

We analyze the colors and sizes of 32 quiescent (UVJ-selected) galaxies with strong Balmer absorption ($\mbox{EW}(H\delta) \geq 4$\AA) at $z\sim0.8$ drawn from DR2 of the LEGA-C survey to test the hypothesis that these galaxies experienced compact, central starbursts before quenching. These recently quenched galaxies, usually referred to as post-starburst galaxies, span a wide range of colors and we find a clear correlation between color and half-light radius, such that bluer galaxies are smaller. 
We build simple toy models to explain this correlation: a normal star-forming disk plus a central, compact starburst component. Bursts with exponential decay timescale of $\sim$~100 Myr that produce $\sim10\%$ to more than 100\% of the pre-existing masses can reproduce the observed correlation. More significant bursts also produce bluer and smaller descendants. Our findings imply that when galaxies shut down star formation rapidly, they generally had experienced compact, starburst events and that the large, observed spread in sizes and colors mostly reflects a variety of burst strengths. Recently quenched galaxies should have younger stellar ages in the centers; multi-wavelength data with high spatial resolution are required to reveal the age gradient. 
Highly dissipative processes should be responsible for this type of formation history. While determining the mechanisms for individual galaxies is challenging, some recently quenched galaxies show signs of gravitational interactions, suggesting that mergers are likely an important mechanism in triggering the rapid shut-down of star-formation activities at $z\sim0.8$.

\end{abstract}


\section{Introduction}

In the last decades, large extragalactic surveys have established that there are two general categories of galaxies; blue, star-forming, morphologically disk galaxies and red, quiescent, predominantly spheroid galaxies \citep{str01,bal04,bel04}. Both the number density and the total stellar mass density of quiescent galaxies increase with time since at least $z\sim4$ \citep{bel04,fab07,ilb13,muz13b}. Since quiescent galaxies have little in-situ star-formation activity, the growth of the number and stellar mass densities indicates that at any cosmic time, part of the star-forming galaxies stop forming stars and join the quiescent population. 

So far, we have not yet understood how and why star formation vanishes. Observations have provided stringent empirical constraints and suggested that there is no simple answer. 
First of all, the stellar populations of galaxies in transition suggest that galaxies spend a wide range of timescales moving from the star-forming to the quiescent population \citep{mart07}. In addition, joint analysis of the structure and star-formation histories of galaxies demonstrated that there should be multiple mechanisms at work. One type of processes shuts off star formation slowly without necessarily changing the structures of galaxies. On the other hand, some galaxies must have experienced rapid transformation in both structure and star-formation rates in the same short period of time \citep{sch14,nog18}. These two types of evolutionary tracks co-exist from the local universe up to at least $z\sim2$ \citep{barr13,wu18a,bel19,nog19}. The relative contribution to the build-up of the quiescent population varies with time such that the rapid processes, so-called quenching, is less important in the local universe but likely contributes a significant fraction of the quiescent population as redshift approaches unity and beyond \citep{gon12,wil16,row18}. 

Recent quenching processes leave unique imprints in stellar populations of galaxies. These galaxies have abundant A-type stars, a sign of star-formation activity $\sim1$~Gyr ago, but lack of hot, ionizing O- and B-type stars. Traditionally, these galaxies are often called post-starburst galaxies \citep{dre83,bal99,dre99}. In this paper, we call them recently quenched galaxies because a starburst, an elevated level of star-formation, is not required to produce such a stellar population \citep{leb06}. We focus on massive ($M_\ast > 10^{10}M_\odot$) recently quenched galaxies at $0.6 < z < 1.0$, a lookback time of $\sim6-8$~Gyrs. Previous studies found that the fraction of recently quenched galaxies in this redshift range is $\gtrsim10$ times higher than in the local universe \citep{yan09,ver10} and they may contribute to a non-negligible fraction of newly-formed quiescent galaxies. The knowledge of quenching is a crucial building block for understanding galaxy evolution at this epoch.

The properties of recently quenched galaxies provide important insights to the quenching mechanisms. Previous studies on the structure and morphology show that at $z\sim1$, recently quenched galaxies have a high incidence of asymmetric structures \citep{ver10} and a minority of them exhibit clear tidal features \citep{wu14}, suggesting that at least part of recently quenched population of galaxies are produced by galaxy mergers or interactions. 

Meanwhile, recently quenched galaxies at $z>0.5$ tend to have smaller effective radii ($R_e$) than the average quiescent galaxies and are also much more compact than normal star-forming galaxies \citep{whi12,bel15,yan16,alm17,wu18a,bel19}. This compactness is considered as evidence that quenching is accompanied by a structural transformation, likely driven by highly dissipative processes that induce a centrally concentrated starburst event right before star-formation stops \citep{wu18a,bel19}.
Nevertheless, it is not clear whether there was a starburst event before quenching and if there was, whether the starburst event can produce the compact recently quenched galaxies hypothesized in previous studies. 

In this paper, we will show that the colors and the effective radii, $R_e$, of recently quenched galaxies are correlated and the correlation can be explained as the evolution after centrally-concentrated starburst events happened in star-forming disk galaxies. We describe the recently quenched galaxy sample, the measurements of galaxy properties, and the stellar population synthesis models in Section~\ref{sec:data}. We present the analysis of stellar populations and structures in Section~\ref{sec:ana}. In Section~\ref{sec:model}, we use toy models to interpret the evolution of colors and structures of recently quenched galaxies. We discuss the implication of our results on the galaxy evolution studies in Section~\ref{sec:dis} and give our final remarks in Section~\ref{sec:sum}.

\section{Data and Analysis}
\label{sec:data}

In this study, we analyze the broadband colors and the effective radii of 32 recently quenched galaxies selected from the DR2 of the LEGA-C spectroscopic survey \citep{vdw16,str18}. The LEGA-C survey obtains rest-frame optical spectra of galaxies at $0.6 \leq z \leq 1.0$. The typical continuum signal-to-noise ratio (S/N) is 20~\AA$^{-1}$. The high S/N and spectral resolution allow for quantifying absorption features precisely \citep{wu18b}, which is crucial for obtaining a clean sample. In this study, we use the spectral line measurements from the LEGA-C DR2 catalog \citep{str18}. 

In addition to the LEGA-C DR2 data, we use stellar population synthesis models to track the color evolution of star-formation histories (SFHs) that produce recently quenched galaxies. 

\subsection{The sample of recently quenched galaxies}
\label{sec:sample}
\begin{figure*}
	\includegraphics[width=0.95\textwidth]{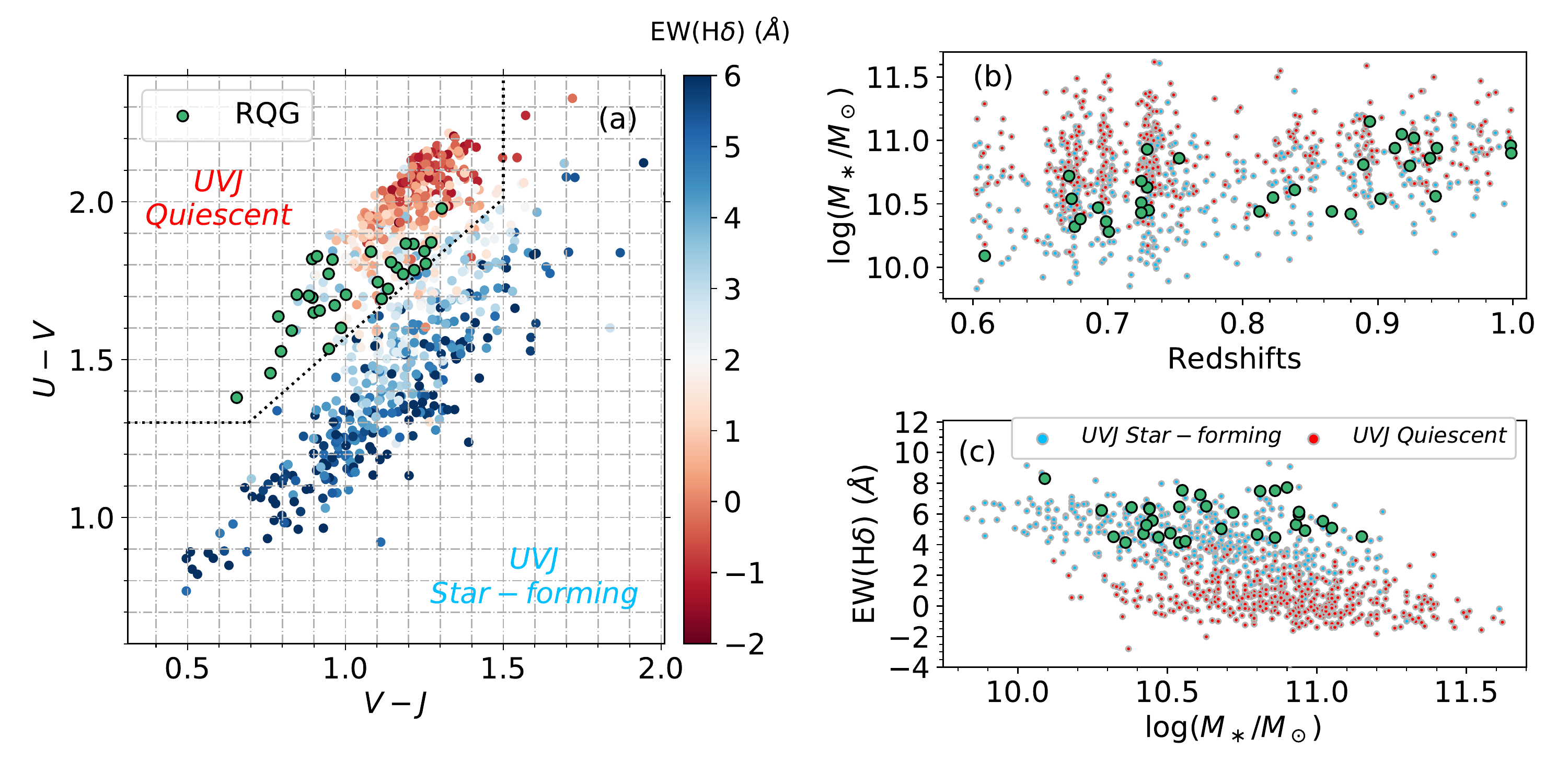}
	\caption{(a) The $U-V$ and the $V-J$ colors of recently quenched galaxies (green circles): quiescent galaxies with $\mbox{EW}(\mbox{H}\delta) \geq 4\AA$. Other galaxies in the LEGA-C DR2 satisfying the quality cut ($\texttt{SN\_RF\_4000} \geq 10$, $\texttt{f\_use} = 1$, $\texttt{f\_int} = 0$, and with EW(H$\delta$) measurement) are also plotted, color-coded by their H$\delta$ absorption strength. The dotted lines are the demarcation separating quiescent galaxies from star-forming galaxies on the UVJ diagram. (b) The stellar masses and redshifts of the recently quenched galaxy sample. (c) The EW(H$\delta$) of recently quenched (green), quiescent, and star-forming galaxies, respectively. The strengths of the EW(H$\delta$) absorption of recently quenched galaxies are comparable to those of star-forming galaxies, indicating young stellar populations.}
	\label{fig:mz_uvj}
\end{figure*}

\begin{figure*}
	\includegraphics[width=0.95\textwidth]{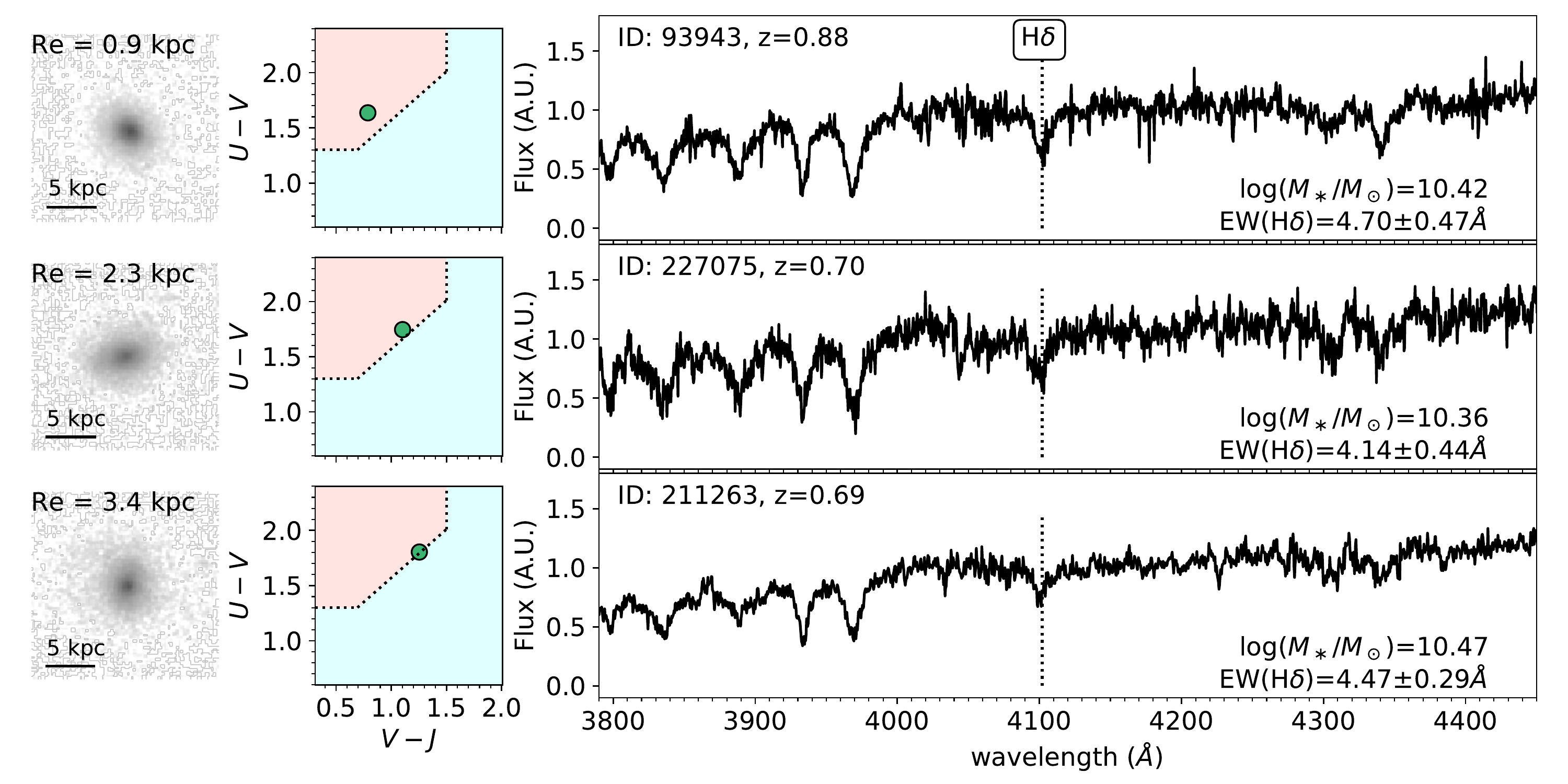}
	\caption{Examples images, colors, and spectra of recently quenched galaxies. \textit{First column:} \textit{HST} F814W images. \textit{Second column:} The rest-frame $U-V$ and $V-J$ colors. \textit{Third column:} The spectra around rest-frame 4000\AA. Recently quenched galaxies in this paper are quiescent galaxies (according to their $U-V$ and $V-J$ colors) with strong Balmer absorption line  ($\mbox{EW}(\mbox{H}\delta) \geq 4$\AA.)}
	\label{fig:spec}
\end{figure*}

\begin{table*}
	\label{tab:sample}
	\begin{center}
		\begin{threeparttable}
			\caption{Sample of Recently Quenched Galaxies}
			\begin{tabular}{ccccccccc}
				\hline
				\hline
				ID & RA & Dec & $\log(M_\ast/M_\odot)$ & Redshift & EW(H$\delta$) (\AA) & U-V & V-J & $R_e$ (kpc)\\
				\hline
				134358 & 150.14000 & 2.37598 & 10.54 & 0.9020 & $4.13\pm0.68$ & 1.87 & 1.22 & 2.73 \\ 
				206858 & 150.09439 & 2.47821 & 10.51 & 0.7248 & $4.74\pm0.21$ & 1.82 & 0.89 & 0.73 \\ 
				211263 & 150.15262 & 2.53226 & 10.47 & 0.6928 & $4.47\pm0.29$ & 1.80 & 1.25 & 3.39 \\ 
				93943 & 150.38770 & 2.32140 & 10.42 & 0.8800 & $4.70\pm0.47$ & 1.64 & 0.79 & 0.87 \\ 
				97310 & 150.41539 & 2.37322 & 10.56 & 0.9428 & $4.22\pm0.77$ & 1.67 & 0.97 & 2.85 \\ 
				133501 & 150.20291 & 2.36306 & 10.93 & 0.7292 & $5.30\pm0.12$ & 1.82 & 0.96 & 3.32 \\ 
				110805 & 150.17149 & 2.08407 & 10.63 & 0.7292 & $6.50\pm0.25$ & 1.78 & 1.22 & 3.41 \\ 
				86349 & 150.41479 & 2.20369 & 10.72 & 0.6713 & $6.09\pm0.19$ & 1.65 & 0.90 & 2.11 \\ 
				88863 & 150.42497 & 2.24202 & 10.44 & 0.8124 & $6.39\pm0.40$ & 1.66 & 0.92 & 1.52 \\ 
				105474 & 150.25063 & 2.01504 & 10.54 & 0.6732 & $6.46\pm0.17$ & 1.53 & 0.80 & 2.34 \\ 
				107468 & 150.36610 & 2.04178 & 11.05 & 0.9178 & $5.08\pm0.23$ & 1.83 & 0.91 & 1.68 \\ 
				127946 & 150.07486 & 2.29866 & 10.86 & 0.9387 & $4.46\pm0.76$ & 1.79 & 1.16 & 4.30 \\ 
				179865 & 149.82149 & 2.44241 & 10.94 & 0.9438 & $5.94\pm0.43$ & 1.87 & 1.19 & 4.19 \\ 
				206499 & 149.90077 & 2.47433 & 10.45 & 0.7303 & $5.57\pm0.30$ & 1.84 & 1.08 & 1.18 \\ 
				113195 & 150.03261 & 2.11639 & 10.96 & 0.9985 & $4.91\pm0.36$ & 1.77 & 1.18 & 3.94 \\ 
				161516 & 149.86713 & 2.23407 & 10.61 & 0.8385 & $7.25\pm0.42$ & 1.98 & 1.30 & 1.16 \\ 
				148698 & 149.77200 & 2.08049 & 10.32 & 0.6755 & $4.51\pm0.25$ & 1.71 & 0.85 & 0.87 \\ 
				216730 & 150.16000 & 2.59545 & 11.15 & 0.8941 & $4.52\pm0.15$ & 1.77 & 0.95 & 3.22 \\ 
				231630 & 150.16969 & 2.78132 & 10.68 & 0.7249 & $5.03\pm0.12$ & 1.70 & 0.89 & 1.65 \\ 
				232099 & 150.15569 & 2.78768 & 10.43 & 0.7248 & $5.27\pm0.21$ & 1.53 & 0.95 & 1.56 \\ 
				203666 & 150.39935 & 2.79416 & 10.55 & 0.8224 & $7.54\pm0.50$ & 1.84 & 1.25 & 2.93 \\ 
				213986 & 149.91373 & 2.56557 & 10.28 & 0.7008 & $6.23\pm0.22$ & 1.59 & 0.83 & 1.55 \\ 
				227075 & 150.07903 & 2.71821 & 10.36 & 0.6988 & $4.14\pm0.44$ & 1.75 & 1.10 & 2.27 \\ 
				230053 & 149.97841 & 2.76018 & 10.38 & 0.6797 & $6.43\pm0.65$ & 1.72 & 1.14 & 2.52 \\ 
				247031 & 149.87163 & 2.60669 & 10.44 & 0.8659 & $6.32\pm0.69$ & 1.81 & 1.14 & 3.75 \\ 
				250391 & 149.87756 & 2.65006 & 10.81 & 0.8893 & $7.49\pm0.16$ & 1.46 & 0.76 & 1.95 \\ 
				260163 & 149.86987 & 2.77514 & 10.86 & 0.7529 & $7.51\pm0.26$ & 1.87 & 1.27 & 4.50 \\ 
				184250 & 150.50551 & 2.50582 & 10.09 & 0.6088 & $8.30\pm0.48$ & 1.69 & 1.11 & 2.43 \\ 
				245150 & 149.75797 & 2.57999 & 10.80 & 0.9239 & $4.65\pm0.57$ & 1.71 & 1.00 & 4.91 \\ 
				246345 & 149.76649 & 2.59571 & 10.90 & 0.9989 & $7.72\pm0.09$ & 1.38 & 0.66 & 1.29 \\ 
				258467 & 149.78896 & 2.75328 & 10.94 & 0.9127 & $6.13\pm0.18$ & 1.60 & 0.99 & 2.04 \\ 
				77161 & 150.67413 & 2.06320 & 11.02 & 0.9268 & $5.53\pm0.24$ & 1.70 & 0.88 & 2.14 \\ 
				\hline
			\end{tabular}
		\end{threeparttable}
	\end{center}
\end{table*}

We are interested in galaxies whose SFR is low but that was much higher in the recent past. Accordingly, our sample must satisfy two criteria: they must have had a rapid drop in star-formation, as evidenced by enhanced Balmer absorption, and they must currently quiescent. 

We use the H$\delta$ absorption as the probe for the recent star-formation activity. We require a fiducial cut of $\mbox{EW}(\mbox{H}\delta) \geq 4$\AA\ after subtracting the contribution from the emission line derived from a joint fitting of the stellar continuum and emission lines using the Penalized Pixel-Fitting (pPXF) method \citep{cap04,cap17}. We select galaxies with S/N higher than 10 per pixel at rest-frame 4000\AA\ ($\texttt{SN\_RF\_4000} \geq 10$) and quality flags $\texttt{f\_use} = 1$ in the LEGA-C DR2 catalog \citep{str18} to minimize mis-classification. The typical uncertainties in EW(H$\delta$) is $\sim0.4$\AA.

There are different methods to select quiescent galaxies. The lack of emission lines associated with star-formation activity are commonly used as a criterion, for example, Balmer lines. However, H$\alpha$ redshifts to the near IR at $z\gtrsim0.5$ and only less than half of the LEGA-C spectra cover both H$\delta$ and H$\beta$ lines. An alternative is the [O\,II]$\lambda3727,3729$ emission. Nevertheless, recently quenched galaxies commonly have detectable [O\,II]$\lambda3727,3729$ emission from ionizing sources other than star-forming regions \citep{yan06}. Requiring the lack of [O\,II] emission leads to an incomplete sample and potential biases against AGN hosts \citep{lem10,wu14,lem17}.

Instead, we use the rest-frame $U-V$ and $V-J$ broadband colors to define quiescence. We derive the color by fitting the observed multiwavelength spectral energy distributions (SEDs) from the UltraVISTA catalog \citep{muz13b} using the FAST code \citep{kri09}. We adopt the demarcation between star-forming and quiescent galaxies proposed by \citet{muz13a}. Recently, selections based on only broadband photometry have been extensively applied \citep[e.g.][]{kri10,whi12}. We will discuss the different methods in Section~\ref{sec:dis}.

We derive $R_e$ of galaxies, the semi-major radius enclosing 50\% of light, using the \textit{HST} ACS F814W images from the COSMOS program \citep[GO-9822, GO-10092, PI: N. Scoville][]{sco07}, following the procedure of \citet{vdw12} using the \texttt{galfit} program \citep{pen10}. Although galaxies do not necessarily have simple S\'{e}rsic profiles \citep[e.g.,][]{mar16}, tests on simulated galaxy images suggest that as long as the S/N ratios of the images are reasonably high ($S/N \gtrsim 100$), $R_e$ can generally be recovered with a $\sim$20\% accuracy even for galaxies with double S\'{e}rsic profiles \citep{dav14,dav16}. While recently quenched galaxies may intrinsically consist of multiple components (Section~\ref{sec:model}), we choose to adopt the $R_e$ from single S\'{e}rsic fits for all galaxies for a consistency in methodology. The \textit{HST} images of all our recently quenched galaxies have $S/N\gtrsim100$, therefore, we expect the $R_e$ is reliable.

We exclude 2 galaxies with catastrophic failures in the \texttt{galfit} fitting or highly disturbed morphologies. One of them exhibits three bright clumps and the other has a diffuse asymmetric low surface brightness feature extending towards a nearby faint object. However, their disturbed morphologies carry important clues for the formation of recently quenched galaxies as will be discussed in Section~\ref{sec:mec}. We also exclude 3 galaxies with close companions that potentially affect broadband photometry thus the estimate of stellar masses and rest-frame colors ($\texttt{f\_int} = 1$). The final sample consists of 32 recently quenched galaxies (Table~\ref{tab:sample}). Fig.~\ref{fig:mz_uvj} shows their redshifts, stellar masses, $U-V$ and $V-J$ colors, and EW(H$\delta$).  Fig.~\ref{fig:spec} shows the \textit{HST} images and spectra of 3 recently quenched galaxies with different colors as examples. Recently quenched galaxies in our sample reside mainly in the average density environment with local overdensity $\log(1+\delta) = 0.1\pm0.4$ (zero is the average) according to the estimate of \citet{dar15}. Our sample thus does not probe extreme environments like galaxy clusters. 

We examine the H$\beta$ emission line strengths of recently quenched galaxies to confirm their quiescence. The spectra of 12 galaxies in the sample cover the H$\beta$ line. Eight of them have $\mbox{EW}(\mbox{H}\beta)_{em} > -1$\AA, consistent with being quiescent \citep{wu18a}. For the remaining 4 galaxies, two galaxies have $\mbox{EW}(\mbox{H}\beta)_{em}$ consistent with $-1$\AA\ (ID: 211263, $-1.20\pm0.23$\AA\ and ID: 230053, $-1.04\pm0.36$\AA), one galaxy (ID: 232099, $-3.92\pm0.45$\AA) has strong [Ne\,III] and [O\,II] emission, indicating an ionizing source other than star formation. One galaxy (ID: 184250, $-2.24\pm0.58$\AA) would be considered as weakly star-forming. Its H$\beta$ emission strength is $\sim0.3$~dex lower than the median value of star-forming galaxies in the LEGA-C DR2 sample ($-4.70$\AA). We conclude that our sample has low SFRs.  

We cross-check the SFH derived from full-spectral fitting technique used by \citet{cha18} and find that our selected galaxies all have an elevated period of star-formation activity in the past $\lesssim1$~Gyr. The strong H$\delta$ absorption feature ensures that these galaxies are truly `recently quenched'. 

\subsection{Stellar population synthesis models}

\begin{figure*}
	\includegraphics[width=0.95\textwidth]{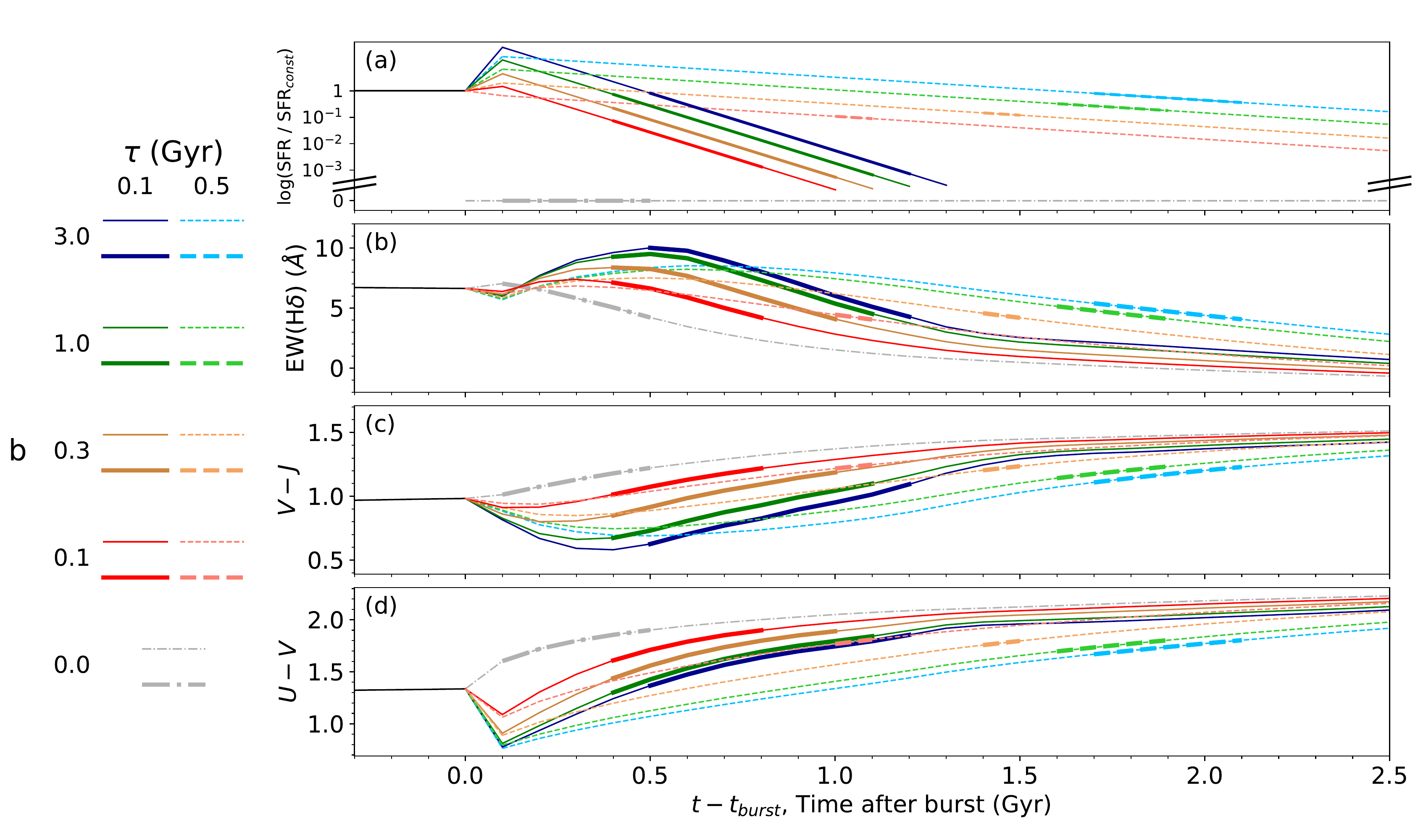}
	\caption{Model star-formation histories and evolution of observables, sampled every 100~Myr. (a) The model star-formation histories. Each model consists of two parts: the constant star formation ($t-t_{burst}\leq0$) and the starburst with  exponential-decay ($t-t_{burst}>0$). The star-formation rates are normalized by the constant star-formation phase. For the starburst phase, the star-formation rates are manually shifted vertically for small amounts for presentation. The parameter $b$ represents the ratio between the stellar masses form in the burst to the constant phase, $b\equiv M_{burst}/M_{const}$, and the $\tau$ is the exponential decay timescale. The $b=0$ represents a pure truncation scenario: no new star form after the constant phase ends. The solid and dashed lines represent for models with $\tau = 0.1$~Gyr and $\tau=0.5$~Gyr, respectively, where the colors distinguish models of different burst strength $b$. The gray dash-dotted line is the model with $b=0$. The truncation at low SFR and changing of y-scale are only for presentation purposes. Thick segmentes show the periods when galaxies are classified as recently quenched galaxies: UVJ quiescent and $\mbox{EW}(H\delta)>4$\AA\ (Section~\ref{sec:sample}). (b) The evolution of equivalent width of H$\delta$ absorption. (c) The evolution of V-J color. (d) The evolution of U-V color.}
	\label{fig:track}
\end{figure*}

Galaxies have a wide range of mass growth and chemical evolutionary histories. Recovering the true formation histories from observables remains a challenging task. In this paper, we use simple star-formation histories combined with stellar population synthesis models to illustrate how recent star-formation events affect observables. We adopt the following analytical form for our fiducial analysis:

%We construct simple analytical forms of star-formation histories that produce recently quenched galaxies and use stellar population synthesis models to track the color evolution.
%%% Old 
%We study 9 different star-formation histories. Eight of them consist of two components: a constant star-formation phase for 4~Gyrs then followed by a subsequent burst with an exponential decay timescale $\tau$ with different strengths of bursts (Fig.~\ref{fig:track}a):

\begin{equation}
\label{eq:sfh}
SFR(t-t_{burst}) \propto \left \{ \begin{array}{ll}
\mbox{constant}, \quad -4\ \mbox{Gyr} \leq t-t_{burst} \leq 0\ \mbox{Gyr}; \\
\exp(-t/\tau), \quad t- t_{burst} > 0\ \mbox{Gyr} .
\end{array}
\right.
\end{equation}

This star-formation history consists of two components: a constant star-formation phase followed by a subsequent burst with an exponential decay timescale $\tau$ with different strength of bursts (Fig.~\ref{fig:track}a).

In this paper, we define the burst strength $b$ as the ratio between the masses formed in the burst to the mass formed in the constant star-formation period, $b\equiv M_{burst}/M_{const}$. We generate 4 sets of models of $b$ = 3, 1, 0.3, and 0.1, with 2 different $\tau$ = 0.1~Gyr and 0.5~Gyr, respectively. 
The combination of $b$ and $\tau$ determines the amount of stars formed right before galaxies become quiescent, which is the key feature for recently quenched galaxies. A shorter $\tau$ and larger $b$ represent a more drastic decline of star-formation rate. A sudden truncation of star-formation corresponds to $\tau=0$. We will see later in Section~\ref{sec:ana} that 0.5~Gyr is roughly the longest $\tau$ that can produce the observation signatures of recently quenched galaxies with our assumed star-formation history. 

The $b=0.1$ model only increases the SFR by $\sim2$ times, averaging over the first 100~Myr after the burst. The $b\geq1$ models would be considered as strong starburst events. More extreme bursts with $b>3$ behave qualitatively similarly to the $b=3$ model for the parameters that we are interested in. We make one more model with $b=0$, mimicking a sudden truncation of star-formation without forming new stars. Fig.~\ref{fig:track}a shows the star-formation histories of the 9 models used in this paper. Real recently quenched galaxies do not have to follow Equation~\ref{eq:sfh}. The parameterization, $b$ and $\tau$, provides insight into how the strength and the timescale of the last period of star-formation affect the color evolution.

We then use the Python implementation of the Flexible Stellar Population Synthesis package \citep[FSPS;][]{con09,con10,for14} to generate the spectra and broadband colors of each model for each 100~Myr time step, adopting the MILES spectral library \citep{san06}, Padova isochrones \citep{gir00,mar07,mar08}, and a Chabrier IMF \citep{cha03}. 

We choose solar metallicity as the fiducial value \citep{gal14}. For the dust attenuation, we use the \citet{car89} dust reddening curve and a 2-component dust model. Young stellar populations ($<10^7$~yr) have $\mbox{A}_{\mbox{v},young}=1.5$ and the rest have $\mbox{A}_{\mbox{v},old}=0.5$. Therefore, when the SFR is low, like the recently quenched galaxies, the total $\mbox{A}_{\mbox{v}}$ of the models approach $\mbox{A}_{\mbox{v}} \simeq 0.5$, which is motivated by the constraints from broadband SED fitting, as well as spectral-photometric fitting of galaxies at similar and slightly higher redshifts \citep{car19b,bel19}. Different $\mbox{A}_{\mbox{v},young}$ makes negligible difference. We use these fiducial values in Section~\ref{sec:ana} and Section~\ref{sec:model} and discuss the assumptions in Section~\ref{sec:dis}.

Fig.~\ref{fig:track} shows the evolution of EW(H$\delta$), $V-J$, and $U-V$ colors of the 9 different star-formation histories with our fiducial choice of metallicity and dust attenuation. The thick segments indicate the period that model galaxies would be classified as recently quenched galaxies based on our selection criteria. 

Galaxies with shorter $\tau$ can be identified as recently quenched galaxies for longer periods of time. We will compare these observables to our recently quenched galaxy sample in Section~\ref{sec:ana} and discuss scenarios that deviate from the fiducial parameters in Section~\ref{sec:model} and Section~\ref{sec:dis}.

\section{Results: Stellar population and structures}
\label{sec:ana}

\begin{figure*}
	\centering
	\includegraphics[width=0.75\textwidth]{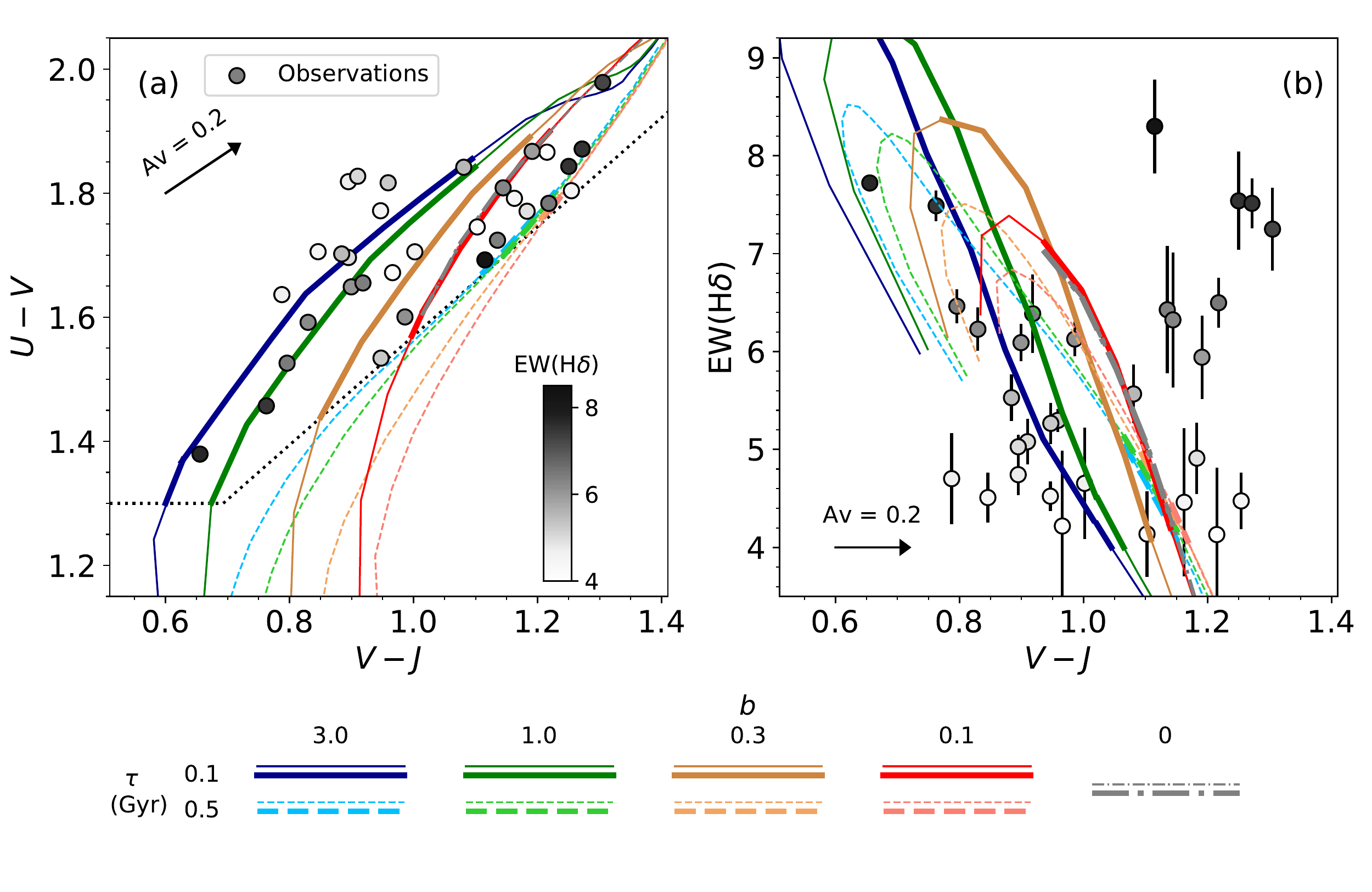}
	\caption{(a): The evolution of $U-V$ and $V-J$ colors under different star-formation histories describe in Section~\ref{sec:data}. The $b=0$ model track largely overlap with the $b=0.1$, $\tau=0.1$~Gyr model (the red line). Circles are the observed recently quenched galaxies, color-coded by their H$\delta$ absorption strength. The model tracks cover distribution of observed recently quenched galaxies on the UVJ color diagram. (b): The EW(H$\delta$) and $V-J$ colors. Models evolve from upper-left to bottom-right. While the majority of recently quenched galaxies follow the model tracks, some galaxies have high EW(H$\delta$) and red colors that are not well explained by the fiducial models. They possibly have larger $A_v$ than assumed.  }
	\label{fig:UVJ_track}
\end{figure*}

We examine the broadband colors and $R_e$ of recently quenched galaxies. We will show that a variety of star-formation histories can reproduce the observed colors. Furthermore, there is a clear correlation between the colors and $R_e$.

\subsection{The rest-frame broadband colors and burst strengths}
\label{sec:cburst}

In general, the recently quenched galaxies have bluer $V-J$ and $U-V$ colors than the rest of quiescent galaxies, reflecting their younger stellar population (Fig.~\ref{fig:mz_uvj}a). On the other hand, they spread over a wide range in the color-color space between $0.7 \lesssim V-J \lesssim 1.3$ and $1.4 \lesssim U-V \lesssim 1.9$.

This sample of recently quenched galaxies cannot be separated based on rest-frame $U-V$ and $V-J$ colors alone. The distributions of colors are much wider than prevalent photometric selections, which usually select UVJ quiescent galaxies with $V-J \lesssim 0.9$ \citep[e.g.][]{whi12,bel19}; only $\sim$30\% of the sample fulfill the color selection. We will discuss the different color distributions in Section~\ref{sec:dis}.

Fig.~\ref{fig:UVJ_track}a shows the recently quenched galaxies and the evolutionary tracks of different star-formation histories on the UVJ diagram. The model galaxies evolve from the bottom-left of the diagram, become redder and move to the top-right. The thick segments indicate the period when the galaxy is classified as a recently quenched galaxy. 

On the UVJ diagram, the models tracks cover the observed distribution of recently quenched galaxies reasonably well. 
The wide range of color can be partially due to different decay time scales $\tau$. For the same burst fraction $b$, recently quenched galaxies with $\tau = 0.1$~Gyr (solid lines) are bluer than galaxies with $\tau = 0.5$~Gyr (dashed lines) in the $V-J$ color.
 
In addition, the $V-J$ color also depends on the burst fraction $b$. Recently quenched galaxies with weaker bursts are on average redder (see also Fig.~\ref{fig:track}). Even with a short decaying timescale of $\tau=0.1$~Gyr, a recently quenched galaxy with $b=0.3$ is hardly ever bluer than $V-J \simeq0.9$. Recently quenched galaxies with even bluer colors are predominantly those experienced strong bursts, which have at least doubled their stellar masses in the starburst events. 

Fig.~\ref{fig:UVJ_track}b shows the EW(H$\delta$) and $V-J$ colors of recently quenched galaxies. In our sample, EW(H$\delta$) ranges from 4\AA\ to $\simeq$8\AA. However, the $\tau=0.5$~Gyr models only produce recently quenched galaxies with $\mbox{EW(H}\delta\mbox{)} \lesssim5$\AA. About half of the sample must have shorter quenching timescales in order to produce the strong H$\delta$ absorption.

The model tracks evolve from the top-left to the bottom-right in Fig.~\ref{fig:UVJ_track}b and are in broad agreement with the majority of the sample, suggesting that our fiducial parameter set is reasonable. However, at the reddest $V-J$ colors, several recently quenched galaxies have high EW(H$\delta$) that are not well explained by the fiducial models. These galaxies possibly have much larger $A_v$ than the fiducial value. We will discuss our assumptions in Section~\ref{sec:dis}.

%several recently quenched galaxies also have both red $V-J$ colors and high EW(H$\delta$) that are not well explained by the fiducial models. These galaxies possibly have $A_v$ much larger than the fiducial value.} We will discuss our assumptions in Section~\ref{sec:dis}. 

\subsection{Colors and $R_e$}

\begin{figure*}
	\includegraphics[width=0.95\textwidth]{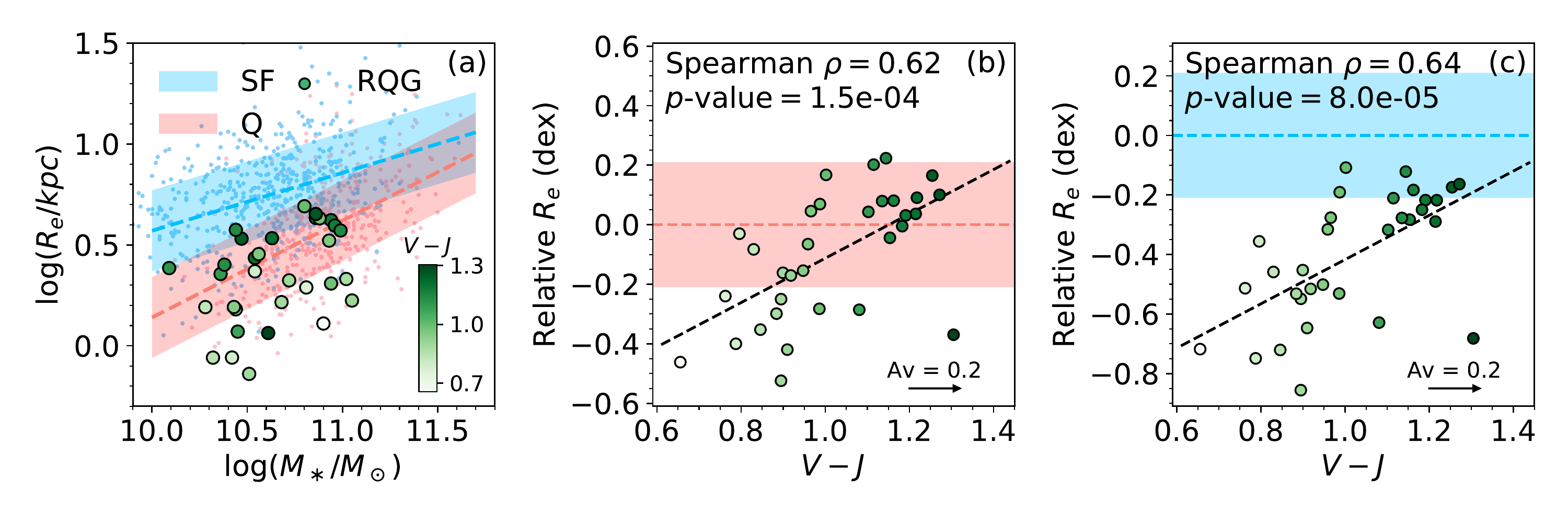}
	\caption{The sizes of recently quenched galaxies. (a) The effective radii, $R_e$, of recently quenched galaxies (green circles) as a function of stellar masses. The blue and red dashed lines are the median mass-size relation for star-forming and quiescent galaxies at $z\sim0.8$ measured from the LEGA-C sample, respectively. Small blue and red dots represent UVJ star-forming and quiescent galaxies in the LEGA-C survey. The shaded regions show $\pm0.2$~dex from the median, which enclose $\sim68\%$ of galaxies in each sample. The color-coding shows the $V-J$ colors of recently quenched galaxies. Bluer recently quenched galaxies have on average smaller $R_e$. (b) The $R_e$ of recently quenched galaxies relative to the median $R_e$ of quiescent galaxies of the same masses (the red dashed line in panel a.) as a function of $V-J$ color. (c) Similar to panel (b), but for the $R_e$ relative to star-forming galaxies. There is a clear correlation between the $V-J$ color and the $R_e$ of recently quenched galaxies.}
	\label{fig:psb_mr}
\end{figure*}

\citet{wu18a} have observed that the $R_e$ of recently quenched galaxies at $z\sim0.8$ are on average smaller than the quiescent population and attributed it to central, compact starbursts happened in the centers of their progenitor galaxies before quenched. This hypothesis leads to a prediction that recently quenched galaxies with stronger bursts should have on average smaller $R_e$. 

As we see in Fig.~\ref{fig:track} and Fig.~\ref{fig:UVJ_track}, the burst strengths correlate with $V-J$ colors. Recently quenched galaxies with blue $V-J$ colors can only be the products of strong starburst events that roughly doubled the stellar masses of galaxies. On the other hand, recently quenched galaxies with red $V-J$ colors can be a combination of galaxies with weak bursts, longer decaying timescale, and are observed later after the starburst. Therefore, we should expect that recently quenched galaxies with blue $V-J$ colors have on average smaller $R_e$. 

Fig.~\ref{fig:psb_mr}a shows the $R_e$ and the stellar masses of recently quenched galaxies, color-coded by their $V-J$ colors. The $R_e$ range from 0.7 kpc to 4.9 kpc, with 16th, 50th, 84th percentiles of 1.2, 2.2, and 4.0 kpc, respectively. We see in general that recently quenched galaxies with blue $V-J$ colors tend to have smaller $R_e$. Next, in order to take into account the intrinsic correlation between the sizes and the stellar masses, we compare the $R_e$ of recently quenched galaxies with only galaxies of similar masses. 

We first fit a linear relation between the median sizes in each 0.1~dex mass bin and the mass in log-log space as:
\begin{equation}
\log(R_e/kpc)_{med} = a \times [\log(M_\ast/M_\odot) - 10.5  ] + b, 
\end{equation}
where (a,b) are ($0.48^{+0.05}_{-0.06}$, $0.38^{+0.01}_{-0.01}$) and ($0.29^{+0.06}_{-0.07}$, $0.71^{+0.01}_{-0.01}$) for quiescent and star-forming galaxies, respectively (dashed lines in Fig.~\ref{fig:psb_mr}a). The 1-$\sigma$ dispersion around the median is $\sim0.2$~dex for both quiescent and star-forming populations in the stellar mass range of interest \citep[also see][]{wu18a}. We then calculate $\Delta \log R_e$, the logarithmic size difference between recently quenched galaxies and the best-fit mass-size relations of quiescent and star-forming galaxies, respectively. We find the $V-J$ colors correlate strongly with $\Delta \log R_e$ when comparing either to quiescent or star-forming galaxies (Fig.~\ref{fig:psb_mr}b,c). Blue recently quenched galaxies are relatively smaller than red recently quenched galaxies. The null hypothesis that the correlation is due to random uncertainty can be rejected with high confidence. The clear correlation between colors and $R_e$ indicates that the star-formation histories and the structures of recently quenched galaxies are coupled.

\section{The size evolution: toy models}
\label{sec:model}

If the progenitors of recently quenched galaxies are drawn from the star-forming population, the correlation between the relative size and $V-J$ color presents a supporting evidence that the formation of recently quenched galaxies is accompanied with `shrinking' of $R_e$, which is a result of centrally-concentrated starburst \citep{wu18a}. 
We build toy models to illustrate how the sizes evolve when centrally-concentrated starbursts happen. 

Based on our model star-formation histories, we assign different structures to each phase of star formation. The constant star-formation components have 2-D mass profiles with a Sersic $n=1$ and $R_e$ ranging from 3~kpc to 8~kpc, corresponding to the sizes of star-forming galaxies of $10^{10} M_\odot \lesssim M_\ast \lesssim 10^{11} M_\odot$. The starburst components have an $R_e$ of 1~kpc and a Sersic $n=4$. The conclusion of this paper are insensitive to the burst light profile for S\'{e}rsic n between 1 and 6.

\begin{figure}[]
	\includegraphics[width=\columnwidth]{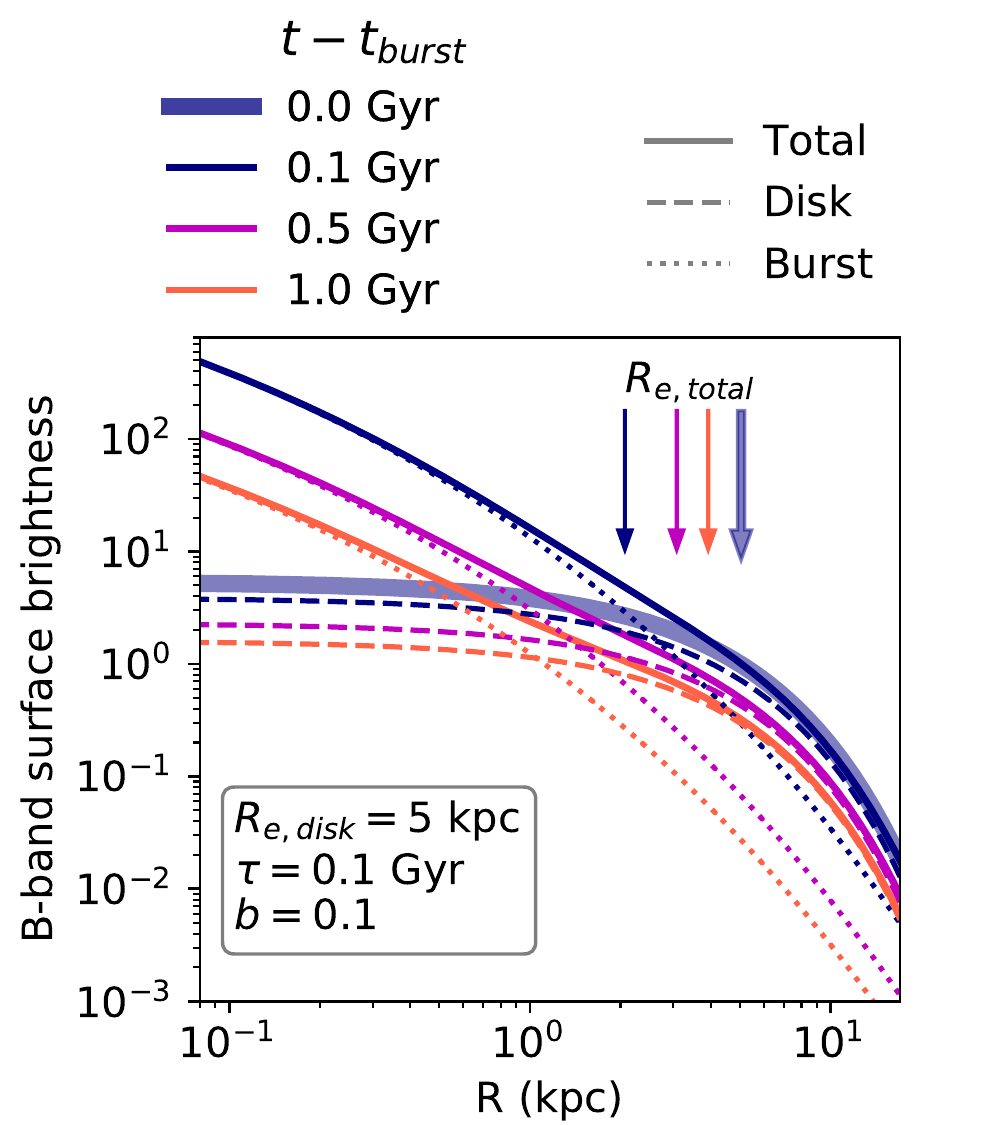}
	\caption{An example of the evolution surface brightness profiles of recently quenched galaxies. The model galaxy has an initial $R_e=5$~kpc for the disk component (thick purple line), followed by a $R_e=1$~kpc burst of $b=0.1$ and $\tau=0.1$~Gyr. The dark blue, magenta, and orange lines represent the profiles at 0.1, 0.5, and 1.0 Gyr after the burst, respectively. Solid lines shows the evolution of the total B-band surface brightness profile. The profiles of the disk and the burst components are dashed and dotted lines, respectively. The light from the burst and the disk component decreases $\sim10$ times and $\sim3$ times in the first 1~Gyr after the burst. The galaxy appears less concentrated as it evolves thus $R_e$ increases. }
	\label{fig:prof_evo}
\end{figure}

The effective radii $R_e$ in this paper is measured in the \textit{HST} F814W filter, close to rest-frame $B$-band at $z\sim0.8$. Therefore, we use the $B$-band mass-to-light ratios generated by the SPS models to convert the mass profiles to the light profiles and measure the $R_e$ in rest-frame $B$-band of the combined light profiles at each time step. Fig.~\ref{fig:prof_evo} shows an example for the evolution of the B-band light profile and $R_e$. 

These toy models do not capture all structure transformation. The transformation of the pre-existing disks, if any, is not taken into account. The change in size also depends on the actual sizes of the starburst components. While this is not well constrained theoretically, starburst cores of $\sim1$~kpc are found in $z\sim2$ compact star-forming galaxies \citep{barr16b} and the smallest $R_e$ of our recently quenched galaxies is also $\sim1$~kpc. We thus choose 1~kpc as the fiducial value. We set the size of the starburst component the same for all models to keep them simple and treat them as illustrative cases.

\begin{figure*}[]
	\includegraphics[width=\textwidth]{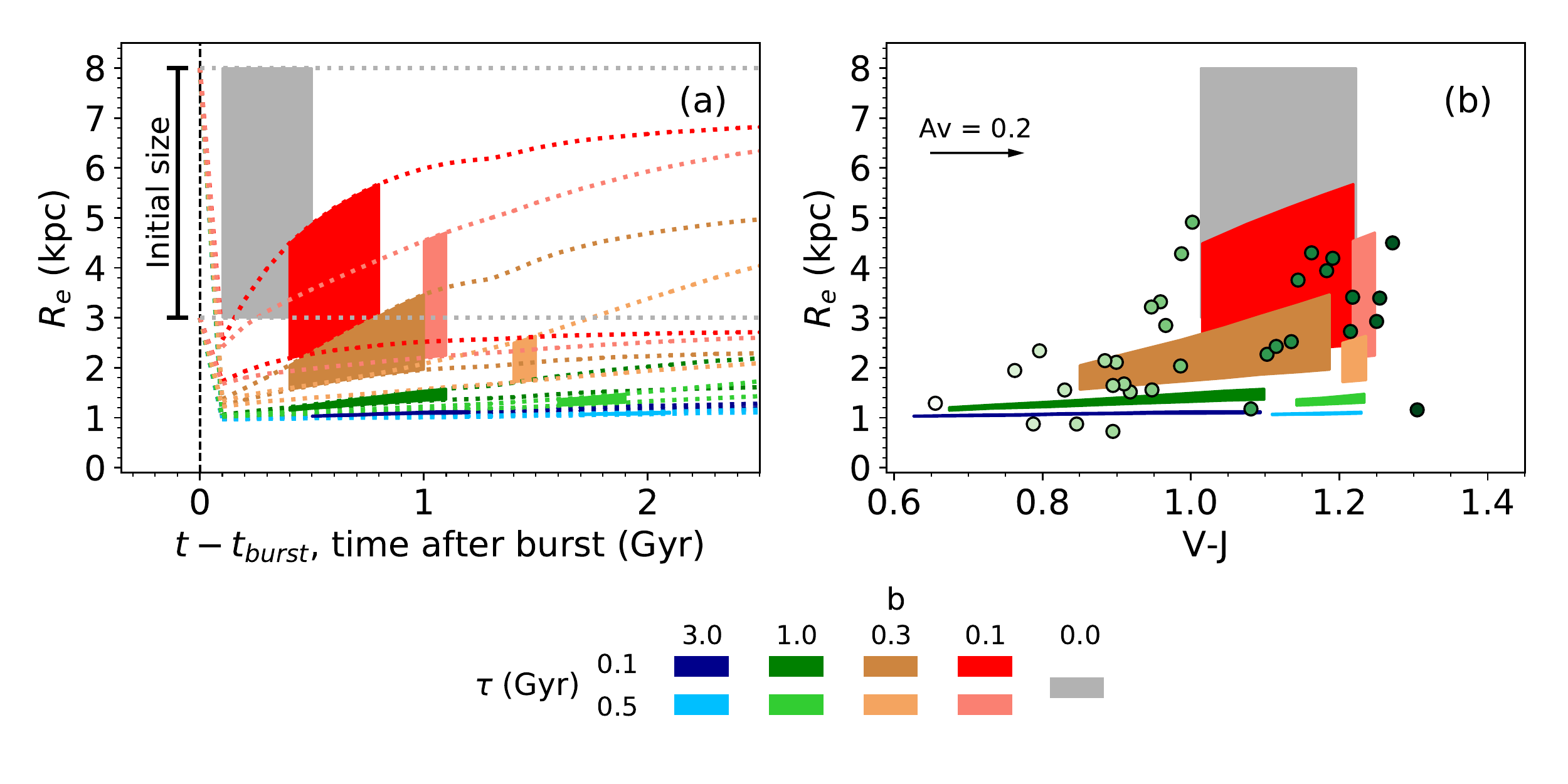}
	\caption{Size evolution of recently quenched galaxies. (a) The $R_e$ evolution as a function of time for the models in Section~\ref{sec:model}. Dotted lines show the upper and lower limits of $R_e$ for each SFH, generated from initial disks of 3 and 8~kpc. The colored areas are the $R_e$ when model galaxies are classified as recently quenched with initial $R_e$ between 3 and 8~kpc. (b) The $R_e$ of recently quenched galaxies as a function of $V-J$ colors. The color areas are the same models as in panel (a). Circles are observed recently quenched galaxies. The arrow indicates the effect of attenuation, assuming it is uniform across the galaxy. The possible ranges of colors and $R_e$ of $\tau=0.1$~Gyr models match the observed correlation.}
	\label{fig:evo}
\end{figure*}

Fig.~\ref{fig:evo}a shows the size evolution of model galaxies with different star-formation histories and initial disk sizes between 3 and 8~kpc. The colored areas highlight the period and the $R_e$ when model galaxies are in the `recently quenched' phases. 

The mass distribution becomes more centrally-concentrated after the burst. Furthermore, the central burst component has a lower mass-to-light ratio, which makes the light distribution even more compact thus the $R_e$ becomes smaller. During the `recently quenched' phase, the M/L of the burst component is $\sim4$ times lower than that of the disk component. Even a weak burst of $b=0.1$ can make a significant difference. The $R_e$ becomes $\sim40\%$, or $\sim0.2$~dex, smaller. This change is roughly the difference between the average star-forming galaxies and those redder recently quenched galaxies (Fig.~\ref{fig:psb_mr}c). For strong bursts of $b\geq1$, the burst component dominates the light output, the sizes of recently quenched galaxies is nearly independent of the sizes of progenitors. If the $R_e$ were measured from the stellar mass distribution, the size evolution would be milder. 

While the mass distribution does not change after the burst event, the $R_e$ still evolves because the M/L ratio evolves. The light emitted from the burst component fades faster than the disk component. Therefore, the light profile becomes less compact and $R_e$ increases (Fig.~\ref{fig:prof_evo}). 

To compare with data, it is more convenient to use the color instead of the time in the model.  Fig.~\ref{fig:evo}b shows the $R_e$ and $V-J$ colors of models and observed recently quenched galaxies. The $R_e$ of model galaxies in recently quenched phases depend mainly on $b$. Models with different $\tau$ produce recently quenched galaxies of very similar $R_e$ but different $V-J$ colors. 

We find that the $\tau=0.1$~Gyr models with $b$ from 0.1 to 3 reproduce the observed correlation reasonably well. On the other hand, the $\tau=0.5$~Gyr models are redder. The $b=0.1$ and $b=0.3$ models are still in agreement with the data but $b\geq1$ models are much redder than most of galaxies for their sizes. 

Overall, the observed correlation can be broadly reproduced by our fiducial toy models: pre-existing star-forming disks plus a centrally-concentrated starburst event with an exponential decay time scale of $\tau = 0.1$~Gyr of various burst strengths ranging from at least $\sim10\%$ to more than $\sim100\%$.

\section{Discussion}
\label{sec:dis}

We have seen that recently quenched galaxies spread over a wide $V-J$ range (Fig.~\ref{fig:mz_uvj}), which suggests a variety of star-formation histories: burst strengths, decay timescales, and ages (Fig.~\ref{fig:UVJ_track}). Meanwhile, the $V-J$ colors correlate strongly with $R_e$ (Fig.~\ref{fig:psb_mr}) and can be explained as a result of concentrated starburst events with various burst strengths (Fig.~\ref{fig:evo}). In this section, we discuss the implications of these results for the formation of recently quenched galaxies and the quenching process. We also discuss the assumptions that we make for the analysis. 

\subsection{Assumptions in the analysis}

\subsubsection{Effect of dust attenuation}

\begin{figure}[]
	\includegraphics[width=\columnwidth]{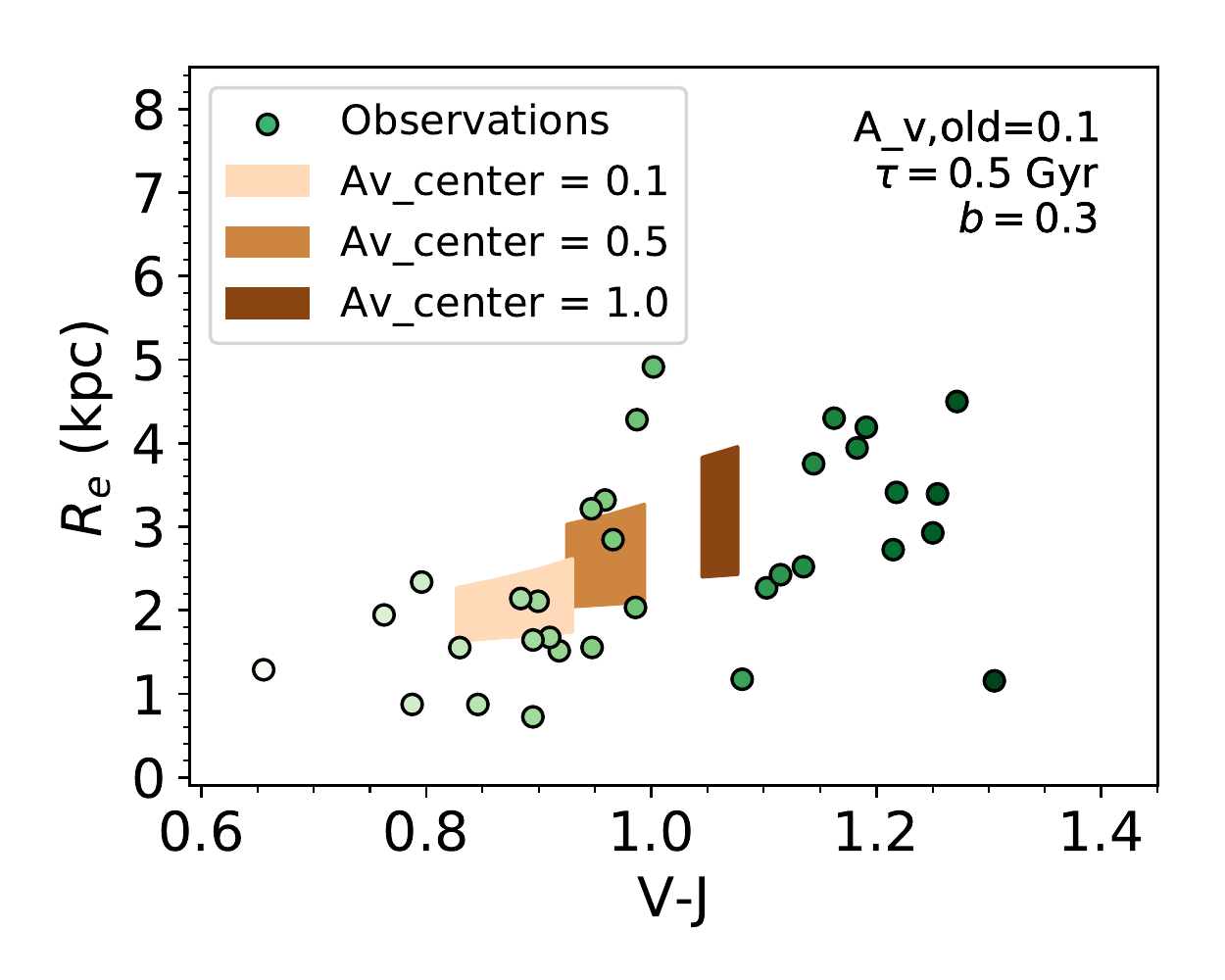}
	\caption{An illustration of the effect of non-uniform dust attenuation on color and size. The 3 model galaxies have the same star-formation history and mass distribution of $\tau=0.5$~Gyr, $b=0.3$, and $\mbox{A}_{\mbox{v},old} = 0.1$. We then artificially impose heavier attenuation at the center 1~kpc for two galaxies, $\mbox{A}_{\mbox{v,center}}=0.5$ and 1.0. Model galaxies with heavy attenuation in the centers become both redder and larger because a larger fraction of light from the compact burst component is attenuated.}
	\label{fig:evo_cen}
\end{figure}

Our models assume an $\mbox{A}_{\mbox{v},old} = 0.5$, which is close to the result of broadband SED fitting, as well as quiescent galaxies at $z \sim 1 - 2.5$ \citep{bel19,car19b}. On the other hand, low-redshift massive ellipticals have on average lower $\mbox{A}_{\mbox{v},old} \simeq 0.15$ \citep{toj13,cit16,fer18}. Assuming a lower $\mbox{A}_{\mbox{v},old}$ brings $\tau=0.5$~Gyr models closer to the observations. We have tested models with different $\mbox{A}_{\mbox{v},old}$ and find that if recently quenched galaxies have $\mbox{A}_{\mbox{v},old}$ ranging from 0.1 to 0.5, the $\tau=0.5$~Gyr models would be able to reproduce the wide range of $V-J$ color. A range of burst strength $b$ is still required to reproduce the distribution of $R_e$. 

Several recently quenched galaxies in our sample have both strong H$\delta$ absorption and red colors (Fig.~\ref{fig:UVJ_track}b), indicating that the $A_v$ of these galaxies is likely much higher. Our assumption of a single $A_v$ value provides a simplified picture that allows us to investigate the interplay of star-formation histories, colors, and structures. However, better estimates on the dust attenuation for individual galaxies will still be needed to improve the constraints on star-formation histories. 
%We are working on deriving the dust attenuation from joint fit of spectroscopy and photometry (Bari\v{s}i\'{c} et al., in prep). A few recently quenched galaxies in our sample have among the strongest H$\delta$ absorption and reddest colors (Fig.~\ref{fig:UVJ_track}), indicating that the $A_v$ is much higher than other recently quenched galaxies. Better constraints on the dust attenuation on individual galaxy basis will improve the constraints on star-formation histories. 

%** Should I add that I did a sanity check: if I plot the "intrinsic" V-J color assuming the photometric Av is correct, the color-size relation in Fig. 5 is still there. Then need to rewrite the previous paragraph e.g.:
%For a sanity check, we calculate the `intrisic' $V-J$ colors by assuming the $A_v$ from SED fitting is correct. We still find a correlation between the color and size. 
%***

Our models also assume that dust attenuation curves of the disk and the burst components have the same dependence on SFR. 
While the shapes of attenuation curves can vary \citep{kon04,kri13,sal18} and the dust property can be different in the burst component \citep{sme18}, exploring its effect on structures over the entire parameter space is beyond the scope of this paper. 

Instead, we show an illustrative case on how spatially non-uniform attenuation may affect the color and size. 
Fig.~\ref{fig:evo_cen} shows three models with $b=0.3$, $\tau=0.5$~Gyr, and  $\mbox{A}_{\mbox{v},old} = 0.1$. For two of them, we artificially increase the $\mbox{A}_{\mbox{v}}$ at the center 1~kpc to 0.5 and 1.0, respectively. With heavier attenuation in the center, model recently quenched galaxies are not only redder but also have larger $R_e$ because a larger fraction of light from the central compact component attenuate, causing the light profile to become less compact. Non-uniform attenuation can qualitatively create the correlation between the color and the size but a large variation will be needed to produce the observed distributions of color and $R_e$ simultaneously. 

\subsubsection{Aperture effect}

In this paper, we are interested in galaxies that are overall quiescent but with strong H$\delta$ absorption showing that they quenched recently. The quiescence is defined by the broadband colors measured in a 2\arcsec.1 aperture that encloses most of the stellar light from galaxies \citep{muz13a}. However, the spectra are measured from slits of a width of 1\arcsec. Part of the light in the photometric aperture falls out of the slits. The EW(H$\delta$) measured from the slit may not be representative for the entire galaxy because the stellar population is not uniform across the entire galaxy and thus could affect the sample selection.

While spectra that match the photometric aperture are not available, we use our slit spectra to investigate how much the gradient of stellar population would affect the sample selection. We select 219 UVJ quiescent galaxies in the LEGA-C survey whose major axes align with the slits ($\Delta P.A. < 45\deg$) so that most of them are at least marginally resolved. We then measure the EW(H$\delta$) from the central 1\arcsec along the slit direction, $\mbox{EW}(\mbox{H}\delta)_{cen}$. So, $\mbox{EW}(\mbox{H}\delta)_{cen}$ is measured from a square aperture of 1\arcsec on a side. We find an average difference between values measured from the center and the entire slit $\mbox{EW}(\mbox{H}\delta)_{cen} - \mbox{EW}(\mbox{H}\delta) = 0.03$\AA\ with a standard deviation of 0.30\AA. The standard deviation is slightly smaller than the typical uncertainty of the EW(H$\delta$) ($\simeq 0.4$\AA). This test suggests that for quiescent galaxies, the effect of population gradient on the EW(H$\delta$) is generally weak. The limited slit size unlikely biases the sample selection.

% To referee:
% The overall comment on the color gradient of quiescent and recently quenched galaxies are in other sections. In this particular section, we focus on what the analysis could be affected. Since the colors are measured from close to the entire galaxies, the color gradient is not an issue. 

\subsection{How galaxies quench?}
\label{sec:mec}

\subsubsection{Should there be a structural transformation?}
The correlation between colors and $R_e$ of recently quenched galaxies suggests a correlation between formation histories and structures. In Section~\ref{sec:model}, we start from an ansatz that the recent quenched galaxies are descendants of normal star-forming galaxies which have experienced central starburst of various burst strengths and quenching timescales. 

An alternative assumption would be that smaller galaxies form in shorter timescales thus have bluer colors and recently quenched galaxies are those extremely compact systems with amongst the shortest formation times and just formed most of their stars recently. 
However, the most compact and fast evolving galaxies should have formed in the early universe, not at $z\sim0.8$, where the universe is already $\sim7$~Gyr old.
Furthermore, not all recently quenched galaxies are compact systems. About 40\% of the recently quenched galaxies have $R_e$ larger than the median mass-size relation of quiescent galaxies (Fig.~\ref{fig:psb_mr}). 
We thus consider the correlation between colors and $R_e$ of recently quenched galaxies in our sample is a result of structural transformation but unlikely the initial condition of the system. 

Under the framework in Section~\ref{sec:model}, the central surface mass density increases sharply during the quenching process. Several studies have pointed out a tight correlation between the quiescence in star formation and the high density in the centers of galaxies \citep{che12,fan13,mos17}. Even a weak burst that adds 10\% of the total stellar mass ($b=0.1$) doubles the surface mass density in the central 1~kpc. A stronger burst of $b=1$ increases the central surface mass density by an order of magnitude. For example, for a S\'{e}rsic $n=1$ disk with $\log(M_\ast/M_\odot) = 10.5$, a $b=1$ compact burst component increases the surface density in the central 1~kpc from $10^{8.7} M_\odot \mbox{kpc}^{-2}$ to $10^{9.7} M_\odot \mbox{kpc}^{-2}$, comparable to the threshold surface density that separates star-forming from quiescent galaxies \citep{che12,mos17}. The central compact burst effectively pushes star-forming galaxies crossing the empirical boundary of central surface density.

\subsubsection{What mechanisms quench galaxies?}

The centrally-concentrated starburst is likely triggered by some violent events that make gas lose angular momentum thus collapses rapidly to the center of galaxies and form an ample amount of stars in a short period of time. Gas-rich mergers have been long hypothesized as a plausible mechanism \citep{hop06,sny11}. A large amount of gas is brought from galaxy outskirt during the merging process and induces not only strong starburst but also AGN activities which consume, heat up, and expel gas in a short time then quenches the galaxy \citep{spr05,hop06}.

Our models favor short burst timescale ($\tau\simeq0.1$~Gyr) and wide range of burst strengths (Section~\ref{sec:model}), which are also found in hydrodynamical merger simulations \citep{dim08}.  To represent the stars formed out of low-metallicity gas brought from galaxy outskirts in the merger scenario  \citep{rup10,kew10}, we have also calculated another set of models, setting the metallicities of the burst components to a lower value of  $\log(Z/Z_\odot) = -0.4$. These low-metallicity settings still reproduce the observed correlation between colors and $R_e$.

The merger scenario can also explain the environmental dependence of the fraction of recently quenched galaxies. At $z\sim1$, recently quenched galaxies are more common in high-density regions even beyond the virial radii of galaxy clusters \citep{wu14}. This is qualitatively consistent with the environmental dependence of the merger rates derived from semi-analytical models \citep{jia12}. However, previous studies at $z\sim1$ found that only a minority of recently quenched galaxies have clear tidal arms or highly disturbed morphologies \citep{wu14}. We find similar behavior in our study. In our sample selection, we find 3 recently quenched galaxies with a close companion, as well as one galaxy with a clumpy structure, and one with an asymmetric extended low surface brightness outskirt. Compared to other 32 recently quenched galaxies that can be reasonably modeled by a S\'{e}rsic profile (see Section~\ref{sec:data}), only $\lesssim15\%$ of recently quenched galaxies are considered highly disturbed or with a companion. 

These recently quenched galaxies with disturbed morphologies provide evidence that galaxy mergers or gravitational interactions contribute at least part of the recently quenched galaxy population. If the mergers is the principal mechanism that rapidly shuts down star formation, the low incidence of disturbed morphologies suggests that the recently quenched phase appears after the tidal features settled. Merger-induced tidal tails around the remnant have faded to a low surface brightness and become difficult to observe \citep{yan04}. This is consistent with predictions from N-body smoothed particle hydrodynamical simulation on gas-rich mergers. The highly disturbed structures last for only $\lesssim 500$~Myr and usually happen before the peak of the starburst \citep{lot10a,lot10b}. 

On the other hand, violent disk instabilities (VDI) may also induce the dissipative process that rapidly form stars in galaxy centers even without a merger event \citep[wet compaction,][]{dek14,tac16a}. \citet{zol15} ran a suite of hydrodynamical simulations and showed that VDI increases the stellar mass densities in the central 1~kpc and decreases $R_e$ and quenching may follow. In observational studies, one candidate galaxy quenched through this process had been found at $z=1.7$ \citep{barr16a}. 

The simulations predict that the process driven by VDI typically takes $0.3-0.4$ Hubble time from the onset of compaction to quiescence, which corresponds to $\sim2.5$~Gyr at $z=0.8$. This timescale is significantly longer than most of our preferred models. For $\tau=0.1$~Gyr and $\tau=0.5$~Gyr models, the elevated star-formation periods last for $\sim0.5$~Gyrs and $\sim1.5$~Gyrs before the galaxies are classified as quiescent. If VDI produces recently quenched galaxies, it likely happens on galaxies with more extreme conditions. For example, very compact galaxies are less stable thus the gas inflow may be more effective, feeding the black hole in a shorter timescale to trigger strong AGN feedback \citep[e.g.,][]{nog19}. Extra information, such as spatially resolved kinematics, will be needed to determine whether VDI can produce recently quenched galaxies at $z<1$ on an individual galaxy basis \citep{zol15}.

\subsection{Are all recently quenched galaxies post-starburst?}

The term \emph{post-starburst galaxy} implies that these quiescent galaxies with strong Balmer absorption had experienced periods of elevated star-formation rates before becoming quiescent. However, such a star-formation history is not required. A sudden quenching without a starburst can also produce quiescent galaxies with strong Balmer absorption \citep{leb06}. 

The sizes and colors together suggest that a pure truncation scenario may contribute to the recently quenched population but cannot account for the full sample. Most recently quenched galaxies are significantly smaller than star-forming galaxies (Fig.~\ref{fig:psb_mr}). If the quenching mechanisms only shut down star-formation without producing new stars in the center, the $R_e$ would be similar to the progenitor star-forming galaxies. In addition, without a newly-formed stellar component, the colors of recently quenched galaxies would be too red for at least half of the sample (the gray area in Fig~\ref{fig:evo}b). We thus conclude that most recently quenched galaxies had at least a low level starburst event before turning into quiescence.

Spatially-resolved information will further verify this scenario. The central compact starburst makes the stellar ages younger in the center. Recently quenched galaxies may appear to have blue centers in multi-wavelength \textit{HST} images if the dust attenuation is not severer in the center. While spectroscopy can provide better estimate of stellar ages, large ground-based telescopes with AO or \textit{JWST} will be needed for good S/N and spatial resolution.

\subsection{Selecting recently quenched galaxies}

Traditionally, the Balmer absorption lines are used to select recently quenched galaxies, or post-starburst galaxies as often called in the literature, because they are one of the key features of A-type stars and thus are sensitive to recent star-formation activities of galaxies. However, beyond the nearby universe, high S/N spectra of galaxies are observationally expansive thus other techniques have been explored. 

In recent years, deep multi-wavelength photometry has become a popular alternative. Regardless of details, these photometric selections either explicitly or implicitly select the bluest quiescent galaxies mainly with rest-frame $V-J \lesssim 0.9$ \citep{kri11,whi12,wil14,yan16,bel19}. However, we have demonstrated in the paper that spectroscopically-identified recently quenched galaxies span a wide range of color: $0.7 \lesssim V-J \lesssim 1.3$ and only $\sim30\%$ (10/32) of the sample fulfill the prevalent color cut.

While the majority of recently quenched galaxies in this paper have $V-J>0.9$, they present only a small fraction of quiescent galaxies at given $U-V$ and $V-J$ colors (Fig.~\ref{fig:mz_uvj}a). The photometric-only selections may not be able to effectively identify these redder recently quenched galaxies without introducing severe contamination to the sample. 
Based on our LEGA-C DR2 mother sample (Fig.~\ref{fig:mz_uvj}a), a more generous color cut of $V-J<1.0$ selects 35 quiescent galaxies and 18 of them have $\mbox{EW}(\mbox{H}\delta) < 4$\AA. If we adopt an even more generous color cut of $V-J<1.3$ and $U-V<1.9$ to include all but one recently quenched galaxies, the contamination reaches $\sim70\%$: 112 out of 143 quiescent galaxies that fulfill the color cut have $\mbox{EW}(\mbox{H}\delta) < 4$\AA. 

On the other hand, by selecting extremely blue quiescent galaxies, the photometric selections can obtain clean samples. The blue colors ensure that most of the stars must have formed recently. Therefore, the star-formation rates have to be high in the recent past. These blue quiescent galaxies also have strong Balmer absorption lines. Among 13 UVJ quiescent galaxies with $V-J \leq 0.9$ in Fig.~\ref{fig:mz_uvj}a, all of them have $\mbox{EW}(\mbox{H}\delta) \geq 3$\AA\ and 10 of them have $\mbox{EW}(\mbox{H}\delta) \geq 4$\AA\ \citep[also see][]{mal18}. Nevertheless, this selection is very incomplete. 

Under the framework of Section~\ref{sec:model}, this incompleteness also introduces a selection bias. The starbursts need to produce nearly comparable stellar masses to the pre-existing masses to make the whole galaxies blue enough to be distinguishable from other quiescent galaxies in the color space. Recently quenched galaxies with $V-J < 0.9$ are predominantly galaxies with strong starbursts ($b\geq1$). Current photometric selections serve as effective alternatives when good spectra are not available, but they may only pick up those had at least doubled their stellar masses in the starburst events. Recently quenched galaxies with weaker bursts may be missing from the sample, even though they can be produced by the same mechanism. This potential bias should be kept in mind when interpreting the results.

\section{Summary and conclusion}
\label{sec:sum}

We examine the colors and the effective radii, $R_e$, of 32 recently quenched galaxies at $0.6 < z < 1.0$ with  $M_\ast > 10^{10} M_\odot$ identified in the DR2 of the LEGA-C survey. The recently quenched galaxies are selected as UVJ quiescent galaxies with strong H$\delta$ absorptions, $\mbox{EW}(\mbox{H}\delta) \geq 4$\AA. 

Recently quenched galaxies have a wide range of colors, $0.7\lesssim V-J \lesssim1.3$, and cannot be cleanly separated from other quiescent galaxies on the UVJ diagram.
We find a strong correlation between the colors and $R_e$ such that bluer recently quenched galaxies have smaller $R_e$. We build simple toy models to explain this correlation: a pre-existing disk plus a central, compact, recent starburst component. The newly-formed burst component makes the galaxy blue and its centrally concentrated distribution leads to smaller $R_e$. 
We find that a burst with $R_e=1$~kpc, an exponential decaying timescale of $\sim0.1$~Gyr, and various burst strengths can reproduce the observed colors and $R_e$. 
The mass of the burst components range from $\sim10\%$ to $>100\%$ of the pre-existing components. 

Our toy models predict an age gradient in recently quenched galaxies such that the stellar population should be younger in the center. Multi-wavelength \textit{HST} images or spectroscopy from AO-assisted ground-based telescopes or \textit{JWST} may be able to reveal the age gradient. 

The central starburst should be triggered by highly dissipative processes. If the violent disk instability is the cause, it likely only happens on galaxies with extreme conditions that make the formation timescale extremely short. On the other hand, this type of formation histories is consistent with the expectation from gas-rich mergers. While most of the recently quenched galaxies do not have highly disturbed morphologies, they likely represent the phase after the structures have settled. Gas-rich merger is likely an important mechanism that triggers rapid shut-down of star-formation activities at $z\sim0.8$.

\acknowledgments
We thank the referee for reading the manuscript and providing further insights. This work is based on observations made with ESO Telescopes at the La Silla Paranal Observatory under programme ID 194-A.2005 (The LEGA-C Public Spectroscopy Survey). This project has received funding from the European Research Council (ERC) under the European Union’s Horizon 2020 research and innovation programme (grant agreement No. 683184). PFW acknowledges the support of the fellowship from the East Asian Core Observatories Association. A.G. acknowledges support by the INAF PRIN-SKA2017 program 1.05.01.88.04.ESKAPE-HI. 
%We thank the referee for the valuable comments. 

\bibliography{PSB_size_burst_v6.1.bbl}

\end{document}